\documentclass[10pt,twocolumn,letterpaper]{article}

\usepackage[pagenumbers]{cvpr} 

\usepackage{multirow} 
\usepackage{bbold} 
\usepackage{amsthm} 
\usepackage{amsmath}
\usepackage{amssymb}
\usepackage{amsfonts}
\usepackage[mathscr]{euscript} 
\usepackage{multirow}
\usepackage{colortbl}   
\usepackage{pifont} 
\usepackage{bbm}

\definecolor{cvprblue}{rgb}{0.21,0.49,0.74}
\usepackage[pagebackref,breaklinks,colorlinks,allcolors=cvprblue]{hyperref}

\title{HKRAG: Holistic Knowledge Retrieval-Augmented Generation \\ Over Visually-Rich Documents}

\author{
  Anyang Tong\textsuperscript{1}\thanks{These authors contributed equally.} \,\,
  Xiang Niu\textsuperscript{1}\footnotemark[1] \,\,
  Zhiping Liu\textsuperscript{1} \,
  Chang Tian\textsuperscript{2} \,
  Yanyan Wei\textsuperscript{1} \,
  Zenglin Shi\textsuperscript{1}\thanks{Corresponding author: zenglin.shi@hfut.edu.cn} \, 
  Meng Wang\textsuperscript{1}\\[4pt]
  \textsuperscript{1}Hefei University of Technology \,\,\,
  \textsuperscript{2}KU Leuven
}

\begin{document}

\newtheorem{thm}{CustomTheorem}
\newtheorem{definition}{Definition}   
\newtheorem{lemma}{CustomLemma}      
\newtheorem{corollary}{CustomCor}    
\newtheorem{example}{CustomEx}       
\newtheorem{proposition}{Proposition}

\maketitle

\def\eg{\textit{e.g.}}
\def\ie{\textit{i.e.}}
\def\Eg{\textit{E.g.}}
\def\etal{\textit{et al. }}
\def\etc{\textit{etc.}}
\newcommand{\dimny}{\mathcal{M}\xspace}
\newcommand{\dimyH}{\mathcal{H}\xspace}
\newcommand{\dimyW}{\mathcal{W}\xspace}
\newcommand{\dimH}{\ensuremath{H}}
\newcommand{\dimW}{\ensuremath{W}}
\newcommand{\dimC}{\ensuremath{C}}
\newcommand{\nStage}{\mathcal{L}\xspace}
\newcommand{\scale}{\mathcal{S}\xspace}
\newcommand{\mypartitletwo}[2][2]{\vspace*{-#1 ex}~\\{\noindent {\bf #2}}}
\newcommand{\mypartitle}[1]{\vspace*{-3ex}~\\{\noindent \underline{\bf #1}}}
\newcommand{\dimn}{\ensuremath{M}}
\newcommand{\apriori}{\textit{a priori}\xspace}
\newcommand{\mapping}{\ensuremath{G}\xspace}
\newcommand{\params}{\ensuremath{\theta}\xspace}
\newcommand{\data}{\ensuremath{X}\xspace}
\newcommand{\SV}{\ensuremath{X}\xspace}
\newcommand{\pro}{\ensuremath{P}\xspace}
\newcommand{\gt}{\ensuremath{G}\xspace}
\newcommand{\npro}{\ensuremath{N}\xspace}
\newcommand{\featSpace}{\ensuremath{\mathrm{\cal X}}\xspace}
\newcommand{\lSpace}{\ensuremath{\mathrm{\cal Y}}\xspace}
\newcommand{\labbb}{\ensuremath{\mathbf{t}}\xspace}
\newcommand{\state}{\ensuremath{z}\xspace}
\newcommand{\nframes}{\ensuremath{T}\xspace}
\newcommand{\kupdate}{\ensuremath{\boldsymbol{\varphi}}\xspace}
\newcommand{\sol}{\ensuremath{\boldsymbol{\beta}}\xspace}
\newcommand{\nsamples}{\ensuremath{N}\xspace}
\newcommand{\Msamp}{\ensuremath{M_{\mathrm{s}}}\xspace}
\newcommand{\nparticles}{\ensuremath{P}\xspace}

\newcommand{\nDepth}{\ensuremath{D_{\mathrm{max}}}\xspace}
\newcommand{\nTrees}{\ensuremath{K}\xspace}
\newcommand{\Xmat}{\ensuremath{\mathbf{X}}\xspace}
\newcommand{\Ymat}{\ensuremath{\mathbf{Y}}\xspace}
\newcommand{\HH}{\ensuremath{\mathbf{H}}\xspace}
\newcommand{\Smat}{\ensuremath{\mathbf{S}}\xspace}
\newcommand{\Dmat}{\ensuremath{\mathbf{D}}\xspace}
\newcommand{\eye}{\ensuremath{\mathbf{e}}\xspace}
\newcommand{\err}{\ensuremath{\boldsymbol{\xi}}\xspace}
\newcommand{\coeff}{\ensuremath{\mathbf{w}}\xspace}
\newcommand{\samp}{\ensuremath{\mathbf{x}}\xspace}
\newcommand{\laby}{\ensuremath{\mathbf{y}}\xspace}
\newcommand{\func}{\ensuremath{\mathbf{g}}\xspace}
\newcommand{\thresh}{\ensuremath{\tau}\xspace}
\newcommand{\treedepth}{\ensuremath{\Gamma_{\mathrm{depth}}}\xspace}
\newcommand{\sampler}{\emph{Sampler}}
\newcommand{\normal}{\ensuremath{\mathrm{\cal N}}\xspace}
\newcommand{\ssvmCost}{\ensuremath{\ell}\xspace}
\newcommand{\Perp}{\perp \! \! \! \perp}
\def\ci{\perp\!\!\!\perp}
\newcommand{\RR}{I\!\!R}
\newcommand{\labl}{\ensuremath{y}\xspace}
\newcommand{\mean}{\ensuremath{\mu}\xspace}

\newcommand{\Mult}{\ensuremath{\mbox{Mult}}\xspace}
\newcommand{\Cat}{\ensuremath{\mbox{Categorical}}\xspace}
\newcommand{\argmin}{\mathop{\mathrm{arg\,min}}}
\newcommand{\argmax}{\mathop{\mathrm{arg\,max}}}
\makeatletter
\newcommand*{\rome}[1]{\expandafter\@slowromancap\romannumeral #1@}
\makeatother

\definecolor{orange}{rgb}{1.0, 0.5, 0.0}

\newcommand{\zl}[1]{\textcolor{red}{[\textbf{ZL}: #1]}}
\newcommand{\ay}[1]{\textcolor{blue}{[\textbf{AY}: #1]}}
\begin{abstract}
Existing multimodal Retrieval-Augmented Generation (RAG) methods for visually rich documents (VRD) are often biased towards retrieving salient knowledge(e.g., prominent text and visual elements), while largely neglecting the critical fine-print knowledge(e.g., small text, contextual details). This limitation leads to incomplete retrieval and compromises the generator's ability to produce accurate and comprehensive answers. To bridge this gap, we propose HKRAG, a new holistic RAG framework designed to explicitly capture and integrate both knowledge types. Our framework features two key components: (1) a Hybrid Masking-based Holistic Retriever that employs explicit masking strategies to separately model salient and fine-print knowledge, ensuring a query-relevant holistic information retrieval; and (2) an Uncertainty-guided Agentic Generator that dynamically assesses the uncertainty of initial answers and actively decides how to integrate the two distinct knowledge streams for optimal response generation. Extensive experiments on open-domain visual question answering benchmarks show that HKRAG consistently outperforms existing methods in both zero-shot and supervised settings, demonstrating the critical importance of holistic knowledge retrieval for VRD understanding.
\end{abstract}    
\section{Introduction}

Retrieval-augmented generation (RAG) has emerged as a leading paradigm to enhance the factual accuracy of large language models (LLMs) by grounding them in external knowledge sources \cite{Tang0GRFC25,DongJL0DW25,fan2024survey}. Traditionally, RAG frameworks operate under the assumption of a purely textual corpus. However, a vast amount of real-world information is encapsulated within visually rich documents (VRDs), such as reports, invoices, and web pages, where meaning is conveyed through an intricate interplay of textual and visual elements (\eg, layout, typography, images) \cite{masry2022chartqa,Open-WikiTable,tanaka2023slidevqa,tanaka2021visualmrc,van2023dude}. Applying text-centric RAG to these multimodal documents creates a significant modality gap, leading to substantial information loss and unreliable generation.

\begin{figure}[t]
\centering
\begin{center}
{\includegraphics[width=0.99\columnwidth]{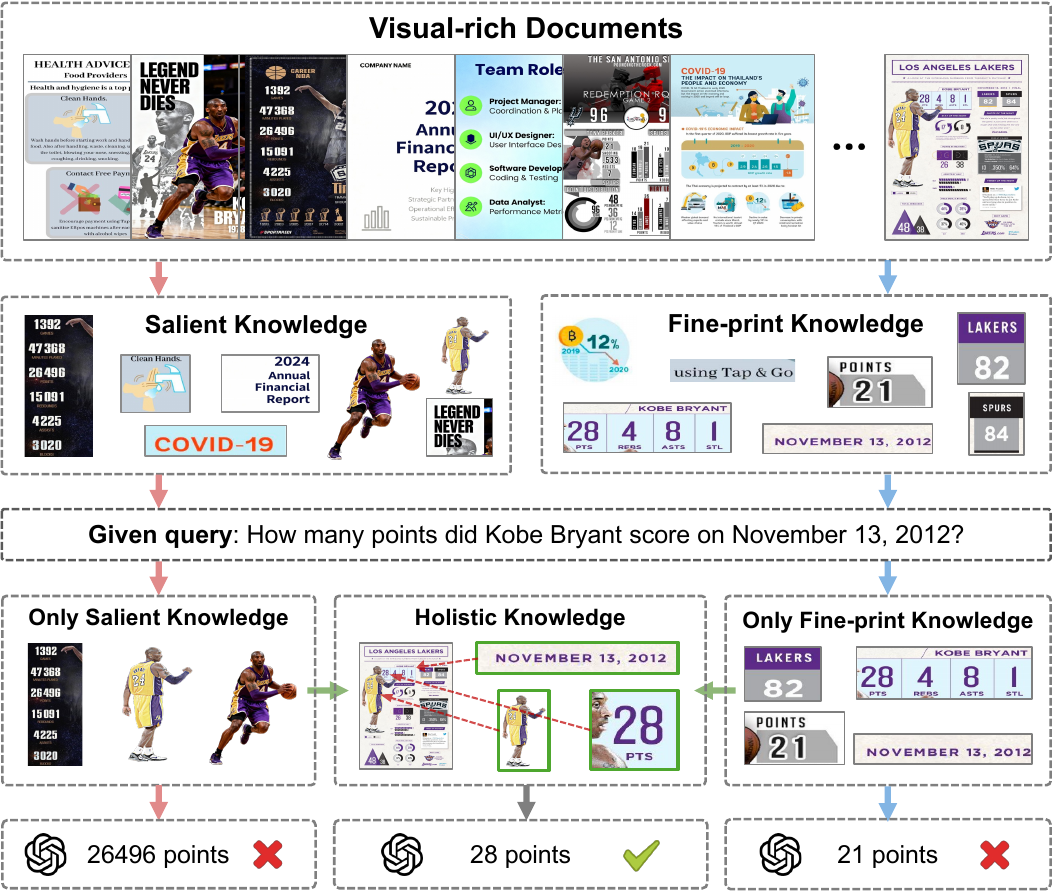}}
\caption{We demonstrate that both salient knowledge and fine-print knowledge are critical for retrieval and generation. Only by jointly leveraging both can we produce reliable answers and effectively mitigate hallucinations.} 
\label{fig:figure1.}
\end{center}
\end{figure}

Recognizing this limitation, several studies \cite{VisRAG,DSE,ColPali} have explored multimodal RAG for document visual question answering (DocumentVQA), typically leveraging large vision-language models (LVLMs) \cite{Tang0GRFC25,DongJL0DW25}. These methods aim to establish a query-to-document alignment by projecting both queries and document images into a shared embedding space. While a step forward, their retrieval mechanism remains fundamentally biased. They are primarily adept at capturing salient knowledge, \ie, information that is visually prominent through large fonts, central placement, or high contrast. Consequently, they often fail to retrieve fine-print knowledge: critical details embedded in small text, footnotes, or dense contextual passages. This bias results in incomplete retrieval, and when the retrieved, saliency-biased context is passed to an LVLM, the generator lacks the necessary context to produce fully accurate answers, especially for queries hinging on nuanced details.

We argue that a comprehensive understanding of VRDs demands holistic knowledge, which necessitates the integration of both salient and fine-print knowledge. The former provides the global context and main points, while the latter contains essential qualifications, conditions, and precise details. However, achieving this is profoundly challenging for two reasons. First, from a retrieval perspective, simultaneously identifying sparse fine-print elements and broad salient features within a complex document is difficult, as standard similarity search tends to be dominated by salient signals. Second, in generation, simply concatenating all retrieved information is suboptimal; the generator must dynamically assess which pieces of knowledge—salient or fine-print—are most relevant to the query and integrate them coherently.

To address these challenges, we propose HKRAG, a novel RAG framework designed for holistic knowledge retrieval and generation over VRDs. Our approach consists of two key components. First, we introduce a \textit{Hybrid Masking-based Holistic Retriever}. It employs explicit masking approaches to separately model and enhance the embeddings for salient and fine-print content within a document, ensuring a query-relevant retrieval process that captures both knowledge types. Second, we develop an \textit{Uncertainty-guided Agentic Generator}. This module does not passively process all retrieved documents. Instead, it first dynamically selects a minimal sufficient set of documents. Then, based on the uncertainty of the initial answers, it adaptively decides how to fuse the complementary information from the salient and fine-print knowledge sources to arrive at a final, well-grounded response.

To summarize, our contributions are as follows:
\begin{itemize}

\item[\normalsize$\bullet$] We identify and formalize a critical limitation in existing multimodal RAG methods for VRDs: their bias towards salient knowledge and their failure to retrieve fine-print knowledge, which leads to the \textit{inadequate holistic-knowledge problem}. 
\item[\normalsize$\bullet$] 
We propose the HKRAG framework, which features (1) a novel hybrid masking-based retriever for balanced knowledge retrieval and (2) an uncertainty-aware agentic generator for dynamic knowledge integration.
\item[\normalsize$\bullet$] 
Extensive experiments on open-domain DocumentVQA benchmarks demonstrate that HKRAG consistently outperforms the existing methods in both zero-shot and supervised settings, validating its effectiveness and generalizability.
\end{itemize}
\begin{figure*}[ht]
\centering
\begin{center}
{\includegraphics[width=0.995\textwidth]{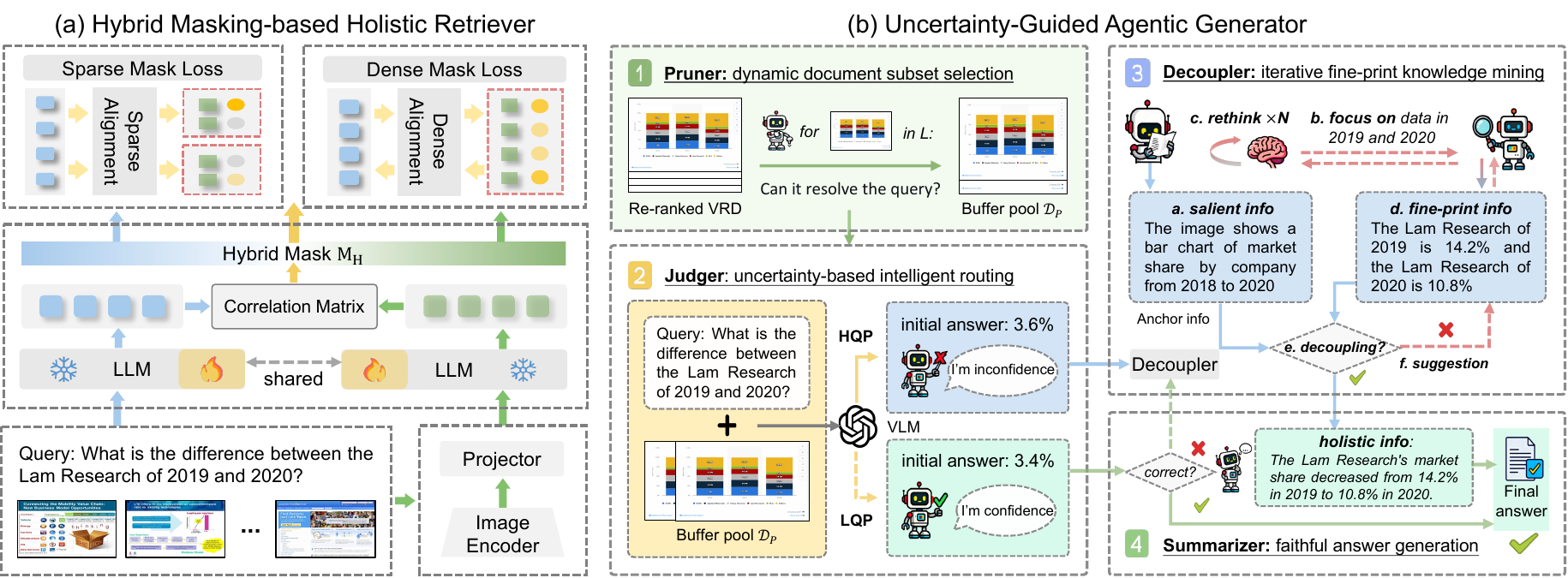}}
\caption{The proposed HKRAG includes (a) Hybrid Masking-based Holistic Retriever and (b) Uncertainty-Guided Agentic Generator.} 
\label{Fig.2.}
\end{center}
\end{figure*}

\section{Related work}
\noindent {\textbf{Retrieval-augmented generation. }}
RAG aims to construct and retrieve external knowledge bases to reduce the generation of hallucinatory content and enhance model credibility across diverse tasks \cite{guo2024lightrag,fan2025minirag,Tang0GRFC25,DongJL0DW25,fan2024survey}. Traditional RAG methods have achieved remarkable success in natural language processing tasks and have been effectively extended to diverse domains such as images, videos, and audio \cite{ren2025videorag,luo2024video}. However, most approaches tend to originate from an academic research perspective, relying on knowledge retrieval from clean corpora \cite{cuconasu2024power,yang2024crag,oettinger1990rag}. In this paper, we focus on retrieving visually rich documents from real-world open-domain scenarios, where answers can only be derived from the documents \cite{masry2022chartqa,Open-WikiTable,tanaka2023slidevqa,tanaka2021visualmrc,van2023dude}. Additionally, the content within these documents is presented in diverse formats, encompassing both prominent and nuanced knowledge that often requires integration to correctly respond to queries. Consequently, establishing an effective RAG pipeline for these documents remains a challenge.

\vspace{0.5em} \noindent {\textbf{RAG for document visual question answering.}}
Visual documents are widely used in real-life scenarios and present content in diverse formats \cite{masry2022chartqa,Open-WikiTable,tanaka2023slidevqa}. To retrieve query-relevant knowledge from visual documents, traditional RAG methods \cite{BM25,Contriever,E5,li2023GTE,E5Mistral,NvEmbed2025} typically rely on text detection techniques (OCR) to scan textual information within documents, severely neglecting the vast amount of visual information present.
Visual large language models (LVLMs) \cite{Tang0GRFC25,DongJL0DW25,xu2024lvlm}, which combine visual encoders with large language models, have gained widespread attention for their image understanding capabilities. Existing RAG methods, \eg, VisRAG \cite{VisRAG}, DSE \cite{DSE}, ColPali \cite{ColPali}, and VDocRAG \cite{tanaka2025vdocrag} for visual documents employ LVLM encoding for both queries and documents, constructing image-based knowledge bases and retrieving knowledge by calculating similarity between external knowledge and queries \cite{VisRAG,DSE,ColPali,tanaka2025vdocrag}. These methods classically leverage query-document alignment to enhance retrieval performance. However, we pinpoint that simultaneously identifying sparse fine-print elements and broad salient features within a complex document is difficult, as standard similarity search tends to be dominated by salient signals. Second, in generation, simply concatenating all retrieved information is suboptimal; the generator must dynamically assess which pieces of knowledge—salient or fine-print—are most relevant to the query and integrate them coherently.
\section{Method}
\label{Method}

\subsection{Problem Formulation}
In the open-domain document visual question answering (DocumentVQA) task, the retriever $\mathcal{R}$ filters query-relevant documents from external visually rich document set and supplies them to the generator $\mathcal{G}$ to improve answer reliability. Both $\mathcal{R}$ and $\mathcal{G}$ are built upon a large vision-language model (LVLM) with a dual-encoder architecture, encoding queries and documents independently.
Specifically, given a textual query $q$ and a collection of $N$ visual documents represented as images $\mathcal{D}=\{d_1,d_2,...,d_N\}$, the retriever is defined as a function $\mathcal{R}:(q,\mathcal{D})\to\mathcal{D}_\mathcal{R}$, which takes $q$ and $\mathcal{D}$ as inputs, and returns a ranked subset $\mathcal{D}_\mathcal{R}=\{d_{r1},d_{r2},...,d_{rk}\} \subset \mathcal{D}$ based on a search algorithm.  $\mathcal{D}_\mathcal{R}$ often consists of the fixed top-$k$ documents most relevant to $q$. For the query $q$ and retrieved results $\mathcal{D}_\mathcal{R}$, the generator can be defined as a function $\mathcal{G}:(q,\mathcal{D}_\mathcal{R})\to a$, which takes $q$ and $\mathcal{D}_\mathcal{R}$ as inputs and concatenates their features to feed into LLM for generating the answer $a$.  

A fundamental limitation of existing multimodal RAG methods in this setting is their inherent bias towards salient knowledge. These methods lack the capability to recognize and retrieve fine-print knowledge. This results in a failure to capture the holistic knowledge necessary for comprehensive document understanding. Consequently, the generator often produces answers based on an incomplete context, leading to factual inaccuracies and hallucinations, especially for queries that require reasoning over subtle details. To overcome this limitation, we propose HKRAG, a framework designed to explicitly model and integrate both knowledge types. It introduces (1) a hybrid masking-based holistic retriever for balanced retrieval of salient and fine-print knowledge, and (2) an uncertainty-guided agentic generator for dynamic, iterative integration of the retrieved knowledge to produce faithful answers.

\subsection{Hybrid Masking-based Holistic Retriever}
To retrieve the query-relevant holistic knowledge from the document set, we first obtain the embeddings of queries and documents. Specifically, each document image undergoes scaling and dynamic cropping processes, which decomposes complex global information into simpler local ones. The resulting document image patches are processed by a visual encoder and subsequently projected into document features through a two-layer MLP. Next, both the query and visual document features are fed into a LLM, from which we extract their final-layer embeddings $(\text{v}_q,\text{v}_d)$, where document embedding $\text{v}_d$ carries substantial information and shares the same dimensions as query embedding $\text{v}_q$.

\noindent {\textbf{\ding{113} Hybrid embedding-masking.}} We then design a hybrid embedding-masking approach that contains a dense mask to localize salient knowledge within query-relevant documents, along with a sparse mask to pinpoint the fine-print knowledge. We begin by normalizing embeddings $(\text{v}_q,\text{v}_d)$ to obtain $\text{v}_q^{n}$ and $\text{v}_d^{n}$, then compute their correlation as $\text{C}_{q,d} = |\text{v}_q^{n} \odot \text{v}_d^{n}|$. To accommodate document features with varying visual information densities and ensure accurate masking, we further scale the correlation values to the range $(0, 1)$, defined by $\text{C}_{q,d}^{n}$: 
\begin{equation}
\text{C}_{q,d}^{n}= \left({1 + \exp\left(-\dfrac{\text{C}_{q,d} - \mu_{\text{C}_{q,d}}}{\sigma_{\text{C}_{q,d}} + \epsilon}\right)}\right)^{-1},
\end{equation} 
where $\epsilon$ is a negligible value, $\mu_{\text{C}_{q,d}}$ and $\sigma_{\text{C}_{q,d}}$ denote the mean and variance of $\text{C}_{q,d}$, respectively. 
Notably, the mean $\mu_{\text{C}_{q,d}^{n}}$ and variance $\sigma_{\text{C}_{q,d}^{n}}$ represent the average level and the fluctuation degree of correlation $\text{C}_{q,d}^n$ between queries and documents, respectively. By adjusting their magnitudes, we define the correlation-based hybrid mask $\text{M}_{\text{H}}$ as follows:
\begin{equation}
\begin{aligned}
\text{M}_{\text{H}} = (&\mathbb{1}(\text{C}_{q,d}^{n} > (\mu_{\text{C}_{q,d}^{n}} - \alpha \cdot \sigma_{\text{C}_{q,d}^{n}}))+ \\
 &\mathbb{1}(\text{C}_{q,d}^{n} > (\mu_{\text{C}_{q,d}^{n}} + \alpha \cdot \sigma_{\text{C}_{q,d}^{n}})))/2,
\end{aligned}
\end{equation}
where $\alpha$ represents the hybrid scaling parameter. The hybrid mask leverages dense masking downward to preserve low-correlation detail features, thereby maintaining embedding diversity and avoiding dominant biases. Simultaneously, it employs sparse masking upward to amplify highly correlated salient contexts, enhancing discernible focal points while suppressing irrelevant noise.

\noindent {\textbf{\ding{113} Bidirectional retriever-tuning.}}
Our goal is not only to bridge the modality gap between textual queries and visual documents, but also to leverage the designed hybrid mask to effectively extract query-relevant holistic knowledge within rich visual documents.
Specifically, we first employ InfoNCE \cite{oord2018representation} as a manner to bridge the modality gap. Positive pairs are formed from query-document pairs $(q,d^{+})$, while in-batch negatives are used to compute the contrastive loss $\mathcal{L}_{\text{IN}}(\text{v}_q, \text{v}_{d^+})$  as follows:
\begin{equation}
\label{Eq1}
\mathcal{L}_{\text{IN}}(\text{v}_q, \text{v}_{d^+}) = -\log \frac{\exp(\text{sim}(\text{v}_q,\text{v}_{d^+}) / \tau)}{\sum_{i=1}^{\mathcal{B}} \exp(\text{sim}(\text{v}_q, \text{v}_{d_i}) / \tau)},
\end{equation}
where $\text{sim}(\cdot,\cdot)$ denotes the cosine similarity function. $\tau$ is a temperature scaling parameter, and $\mathcal{B}$ represents the batch size. 
To achieve more accurate query-to-document alignment, we employ the hybrid mask to compute the dense matching loss $\mathcal{L}_{\text{DIN}}$:
\begin{equation}
\mathcal{L}_{\text{DIN}} = \frac{1}{2}(\mathcal{L}_{\text{IN}}(\text{v}_q, \text{v}_{d^{+}} \cdot \text{M}_{\text{H}})+\mathcal{L}_{\text{IN}}(\text{v}_{d^{+}} \cdot \text{M}_{\text{H}},\text{v}_q)).
\end{equation}
We enhance the alignment between queries and documents in the embedding space, which facilitates the extraction of salient knowledge. To further exploit fine-print knowledge, we utilize $\mathcal{N}$ partial masks to compute the sparse matching loss $\mathcal{L}_{\text{SIN}}$ as follows:
\begin{equation}
\label{Eq5}
\mathcal{L}_{\text{SIN}} = \frac{1}{\mathcal{N}} \sum_{i=1}^{\mathcal{N}} \mathcal{L}_{\text{IN}}(\text{v}_q, \text{v}_{d^{+}} \cdot \text{M}_{\text{H}}^{i}),
\end{equation}
where $\text{M}_{\text{H}}^{i}$ and $\text{M}_{\text{H}}^{i+1}$ have the same dimensions, $\text{M}_{\text{H}}^{i}\cap\text{M}_{\text{H}}^{i+1}=0$, and $\text{M}_{\text{H}}=\bigcup_{N}^{i=1}\text{M}_{\text{H}}^{i} $. The mask partitioning positions are randomized to ensure alignment of details from diverse regions. We employ the loss $\mathcal{L}=\mathcal{L}_{\text{DIN}}+\beta\cdot\mathcal{L}_{\text{SIN}}$ to fine-tune the retriever, where $\beta$ adjusts the contribution ratio between $\mathcal{L}_{\text{DIN}}$ and $\mathcal{L}_{\text{SIN}}$. The retriever learns to map queries and relevant documents into closer proximity within the embedding space, effectively amplifying its sensitivity to query-relevant holistic knowledge. 

\subsection{Uncertainty-Guided Agentic Generator}

Merely concatenating all retrieved information can overwhelm the generator and introduce noise. To enable a more intelligent integration of the retrieved holistic knowledge (encompassing both salient and fine-print knowledge), we introduce an uncertainty-based criterion to classify query-document pairs. Specifically, we define low-uncertainty query-document pair (LQP) and high-uncertainty query-document pair (HQP) below.

\begin{definition}[Low-uncertainty Query-document Pair, LQP]\label{Low-Uncertainty Query-Document Pair}
Given a query $q$ and retrieved documents $\mathcal{D}_\mathcal{R}$, we feed them into LVLM $\Theta$ to obtain an initial answer $\hat{a}=\Theta(q,\mathcal{D}_\mathcal{R})$ and its token sequence $\hat{w}=\{\hat{w}_1,\hat{w}_2,...,\hat{w}_L\}$. We further calculate the average entropy $\mathcal{H}(\hat{w})=- \frac{1}{L} {\textstyle \sum_{i}^{L}}(\hat{w}_i\cdot log(\hat{w}_i))$ of $\hat{w}$ and normalize it to obtain $\mathcal{H}^{'}(\hat{w})\in(0,1)$. When $\mathcal{H}^{'}(\hat{w})$ is lower than the uncertainty threshold $h$, we define this pair as a low-uncertainty query-document pair.
\end{definition}

\begin{definition}[High-uncertainty query-document pair,HQP]\label{High-Uncertainty Query-Document Pair}
Given a query $q$ and retrieved documents $\mathcal{D}_\mathcal{R}$, we compute $\mathcal{H}^{'}(\hat{w})$ according to Definition \ref{Low-Uncertainty Query-Document Pair}. When $\mathcal{H}^{'}(\hat{w})$ is higher than the uncertainty threshold $h$ at time $t$, we define this pair as a high-uncertainty query-document pair at time $t$.
\end{definition}

An LQP indicates that the LVLM can derive sufficient query-relevant knowledge from the provided documents to produce a confident response, no need of further processing. In contrast, an HQP suggests that the initial understanding is inadequate, often due to the need to reconcile subtle relationships between salient and fine-print knowledge. For HQP, we introduce iterative, time-aware reasoning to facilitate deeper knowledge extraction via test-time computing. Building on this classification, we design an uncertainty-guided agentic generator composed of four specialized agents: \textit{Pruner}, \textit{Judger}, \textit{Decoupler}, and \textit{Summarizer}. The overall architecture and interaction flow of these agents are illustrated in Figure \ref{Fig.2.}.


\noindent {\textbf{\ding{113} Pruner: dynamic document subset selection.}}
Existing methods often employ fixed top-$k$ document rankings empirically for generation. While higher $k$ values increase the probability of retrieving query-relevant knowledge, they simultaneously introduce substantial irrelevant noise, particularly when processing visually rich documents, where abundant prominent information obscures query-relevant details. 
To address this issue, we introduce a pruner that evaluates total query-relevant knowledge $\mathcal{I}(q,n)$ of top-$n$ retrieved documents based on LVLM $\Theta$, defined as
\begin{equation}
\mathcal{I}(q,n):= {\textstyle \sum_{i}^{n}} \Theta (q,d_i), d_i \in \mathcal{D}_\mathcal{P}
\end{equation}
where $d_i$ is stored in a dynamic buffer pool $\mathcal{D}_\mathcal{P}$ with capacity $k$ ($n<k$). When $\mathcal{I}(q,n)$ can answer the query $q$, the evaluation process terminates; otherwise, the buffer retains the top-$k$ documents. Finally, $\mathcal{I}(q,n)$ is next passed to the \textit{Judger}.

\renewcommand{\arraystretch}{1}
\begin{table}[t]
\centering
\renewcommand{\arraystretch}{1.1}{
\setlength{\tabcolsep}{1.3mm}{
\begin{center}
\begin{small}
\begin{tabular}{l|cccc}
\hline
\textbf{Datasets} & \textbf{Documents} & \textbf{\#Images} & \textbf{\#Train} & \textbf{\#Test} \\
\hline
\textbf{ChartQA} \cite{masry2022chartqa}& Chart & 20,882 & -- & 150 \\
\textbf{OpenWikiTable} \cite{Open-WikiTable} & Table & 1,257 & 4,261 & -- \\
\textbf{SlideVQA}$\ddagger$ \cite{tanaka2023slidevqa}& Slide & 52,380 & -- & 760 \\
\textbf{VisualMRC} \cite{tanaka2021visualmrc}& Webpage & 10,229 & 6,126 & -- \\
\textbf{InfoVQA} \cite{mathew2022infographicvqa}& Infographic & 5,485 & 9,592 & 1,048 \\
\textbf{DocVQA} \cite{mathew2021docvqa}& Industry & 12,767 & 6,382 & -- \\
\textbf{DUDE} \cite{van2023dude}& Open & 27,955 & 2,135 & 496 \\
\textbf{MHDocVQA}$\ddagger$ & Open & 28,550 & 9,470 & -- \\
\hline
\end{tabular}
\caption{We evaluate our models under both zero-shot and supervised settings. The \textit{zero-shot} evaluation measures their ability to generalize to unseen datasets, whereas the \textit{supervised} evaluation quantifies performance when training data is provided. $\ddagger$ denotes datasets requiring multi-hop reasoning.}
\label{tab:table1}
\end{small}
\end{center}}}
\end{table}
\renewcommand{\arraystretch}{1}

\noindent {\textbf{\ding{113} Judger: uncertainty-based intelligent routing.}}
Beyond ensuring that sufficient knowledge $\mathcal{I}(q,n)$ is retrieved, assessing the reliability of the generated answer is more important. To address this, we introduce a \textit{Judger} that evaluates whether the generator can confidently answer a given query. Specifically, the \textit{Judger} categorizes each query–document pair according to Equations \ref{Low-Uncertainty Query-Document Pair} and \ref{High-Uncertainty Query-Document Pair}. Queries with low uncertainty are deemed straightforward and sent directly to the final agent, the Summarizer, for efficient processing. In contrast, queries with high uncertainty, indicating a potential need to reconcile subtle relationships between salient and fine-print knowledge, are routed to the \textit{Decoupler} for deeper analysis. Notably, the initial answer produced during uncertainty measurement is neither used as model input nor reflected upon, preventing historical outputs from biasing subsequent decisions.

\noindent {\textbf{\ding{113} Decoupler: iterative fine-print knowledge mining.}}
We further introduce a \textit{Decoupler} to deal with the high-entropy queries passed from the \textit{Judger}. The query-relevant salient knowledge is denoted as $\mathcal{I}_h(q,n)$. This information often shares the semantic similarity with the query $q$ but fails to accurately answer it. Therefore, we perform iterative, fine-grained reasoning to mine overlooked fine-print knowledge $\mathcal{I}_d(q,n,t)$ by anchoring on the readily available salient knowledge $\mathcal{I}_h(q,n)$ at time step $t$:
\begin{equation}
\mathcal{I}_d(q,n,t) = \Theta(\mathcal{I}_h(q,n), \mathcal{I}_d(q,n,t-1)).
\end{equation}
To prevent the model from conflating two knowledge types and thereby stimulating the discovery of crucial details that are initially missed, we explicitly decouple them:
\begin{equation}
\mathcal{I}_d^{'}(q,n,t) = \Theta(\mathcal{I}_h(q,n), \mathcal{I}_d(q,n,t)).
\end{equation}
When the explored fine-print knowledge $\mathcal{I}_d^{'}(q,n,t)$ can answer the query or reach the maximum iteration count; we feed both the decoupled salient information and detail information to the \textit{Summarizer}.

\begin{table*}[t]
\centering
\renewcommand{\arraystretch}{1.05}{
\setlength{\tabcolsep}{1.1mm}{
\begin{center}
\begin{small}
\begin{tabular}{l|ccccc|cc|cc|cc|cc} \hline
\textbf{\multirow{2}{*}{Model}} & \multirow{2}{*}{\textbf{Initialization}} & \multirow{2}{*}{\textbf{Docs}} & \multirow{2}{*}{\textbf{Scale}} & \multirow{2}{*}{\textbf{\#PT}} & \multirow{2}{*}{\textbf{\#FT}} & \multicolumn{2}{c}{\textbf{ChartQA}} & \multicolumn{2}{|c}{\textbf{SlideVQA}} & \multicolumn{2}{|c}{\textbf{InfoVQA}} & \multicolumn{2}{|c}{\textbf{DUDE}} \\
 &  &  &  &  &  & \textbf{Single} & \textbf{All} & \textbf{Single} & \textbf{All} & \textbf{Single} & \textbf{All} & \textbf{Single} & \textbf{All} \\
\hline
\multicolumn{13}{c}{\textit{Off-the-shelf}} \\
\hline
BM25 \cite{BM25} & -- & Text & 0 & 0 & 0 & 54.8 & 15.6 & 40.7 & 38.7 & 50.2 & 31.3 & 57.2 & 47.5 \\
Contriever \cite{Contriever} & BERT \cite{BERT} & Text & 110M & 1B & 500K & 66.9 & 59.3 & 50.8 & 46.5 & 42.5 & 21.0 & 40.6 & 29.7 \\
E5 \cite{E5} & BERT \cite{BERT} & Text & 110M & 270M & 1M & 74.9 & 66.3 & 53.6 & 49.6 & 49.2 & 26.9 & 45.0 & 38.9 \\
GTE \cite{li2023GTE} & BERT \cite{BERT} & Text & 110M & 788M & 3M & 72.8 & 64.7 & 55.4 & 49.1 & 51.3 & 32.5 & 42.4 & 36.0 \\
E5-Mistral \cite{E5Mistral} & Mistral \cite{jiang2023mistral7b} & Text & 7.1B & 0 & 1.85M & 72.3 & 70.0 & 63.8 & 57.6 & 60.3 & 33.9 & 52.2 & 45.2 \\
NV-Embed-v2 \cite{NvEmbed2025} & Mistral \cite{jiang2023mistral7b} & Text & 7.9B & 0 & 2.46M & 75.3 & 70.7 & 61.7 & 58.1 & 56.5 & 34.2 & 43.0 & 38.6 \\
CLIP \cite{CLIP2021} & Scratch & Image & 428M & 400M & 0 & 54.6 & 38.6 & 38.1 & 29.7 & 45.3 & 20.6 & 23.2 & 17.6 \\
DSE \cite{DSE} & Phi3V \cite{Phi-3} & Image & 4.2B & 0 & 5.61M & 72.7 & 68.5 & 73.0 & 67.2 & 67.4 & 49.6 & 55.5 & 47.7 \\
VisRAG-Ret \cite{VisRAG} & MiniCPM-V \cite{MiniCPM-V} & Image & 3.4B & 0 & 240K & {87.2*} & 75.5* & 74.3* & 68.4* & 71.9* & 51.7* & 56.4 & 44.5 \\
\hline
\multicolumn{13}{c}{\textit{Trained on OpenDocVQA}} \\
\hline
Phi3 \cite{Phi-3} & Phi3V \cite{Phi-3} & Text & 4B & 0 & 41K & 72.5 & 65.3 & 53.3 & 48.4 & 53.2* & 33.0* & 40.5* & 32.0* \\
VDocRetriever $\dagger$ \cite{tanaka2025vdocrag} & Phi3V \cite{Phi-3} & Image & 4.2B & 0 & 41K & 80.8 & {71.2} & {66.0} & {60.4} & {63.9}* & {48.3}* & {44.1}* & {35.6}* \\
VDocRetriever $\dagger$ \cite{tanaka2025vdocrag} & Phi3V \cite{Phi-3} & Image & 4.2B & 500K & 41K & 84.4 & {76.3} & {75.3} & {69.9} & {70.4}* & {53.9}* & {55.6}* & {47.9}* \\
HKRAG-Retriever & Phi3V \cite{Phi-3} & Image & 4.2B & 0 & 41K & 83.7 & {77.0} & {69.5} & {62.4} & {65.8}* & {52.1}* & {45.4}* & {40.6}* \\
HKRAG-Retriever & Phi3V \cite{Phi-3} & Image & 4.2B & 500K & 41K & \textbf{87.6} & \textbf{82.1} & \textbf{76.2} & \textbf{72.5} & \textbf{72.4}* & \textbf{57.8}* & \textbf{58.9}* & \textbf{51.7}* \\
\hline
\end{tabular} 
\end{small}
\end{center}}}
\vskip -0.1in
\caption{
Comparison of HKRAG’s retrieval performance against embedding-based models and existing approaches in both single-pool and all-pool settings across four open-domain DocumentVQA benchmarks. FT and PT indicate finetuning and pretraining, respectively. * indicates performance on test data for which corresponding training samples are available, and $\dagger$ denotes reproduced results.
\label{tab:table2}}
\end{table*}

\renewcommand{\arraystretch}{1}
\begin{table*}[t]
\centering
\renewcommand{\arraystretch}{1.1}{
\setlength{\tabcolsep}{2.5mm}{
\begin{center}
\begin{small}
\begin{tabular}{c|cc|cc|cc|cc|cc}
\hline
\multirow{2}{*}{\textbf{Generator}} & \multirow{2}{*}{\textbf{Retriever}} & \multirow{2}{*}{\textbf{Docs}} & \multicolumn{2}{c|}{\textbf{ChartQA}} & \multicolumn{2}{c|}{\textbf{SlideVQA}} & \multicolumn{2}{c|}{\textbf{InfoVQA}} & \multicolumn{2}{c}{\textbf{DUDE}} \\
 &  &  & \textbf{Single} & \textbf{All} & \textbf{Single} & \textbf{All} & \textbf{Single} & \textbf{All} & \textbf{Single} & \textbf{All} \\
\hline
Phi3 \cite{Phi-3} & -- & -- & 12.7 & 12.7 & 23.7 & 23.7 & 18.1* & 18.1* & 8.9* & 8.9* \\ \hline
Phi3 \cite{Phi-3} & Phi3 \cite{Phi-3} & Text & 17.7 & 17.5 & 32.4 & 32.1 & 21.0* & 20.3* & 15.4* & 13.7* \\
VDocGenerator \cite{tanaka2025vdocrag} & VDocRetriever \cite{tanaka2025vdocrag} & Image & {54.0} & {51.3} & {52.2} & {53.9} & {39.4}* & {36.2}* & {37.3}* & {36.9}* \\
HKRAG-Generator & HKRAG-Retriever & Image & \textbf{60.7} & \textbf{58.0} & \textbf{56.5} & \textbf{56.6} & \textbf{42.7}* & \textbf{39.6}* & \textbf{45.2}* & \textbf{42.7}* \\
\hline
Phi3 \cite{Phi-3} & Gold & Text & 23.2 & 23.2 & 33.4 & 33.4 & 23.7* & 23.7* & 21.5* & 21.5* \\
VDocGenerator \cite{tanaka2025vdocrag} & Gold & Image & {77.3} & {77.3} & {72.1} & {72.1} & {54.1}* & {54.1}* & {64.5}* & {64.5}* \\
HKRAG-Generator & Gold & Image & \textbf{78.7} & \textbf{78.7} & \textbf{74.0} & \textbf{74.0} & \textbf{54.7}* & \textbf{54.7}* & \textbf{68.8}* & \textbf{68.8}* \\
\hline
\end{tabular}
\caption{DocumentVQA results. All models are fine-tuned on OpenDocVQA. The results marked with * denote performance on unseen test samples, and the other results represent zero-shot performance. The performance gain in green is compared to the text-based RAG that has the same base LLM. Gold knows the ground-truth documents. Models answer the question based on the top three retrieval results.
\label{tab:table3}}
\end{small}
\end{center}}}
\end{table*}
\renewcommand{\arraystretch}{1}

\noindent {\textbf{\ding{113} Summarizer: faithful answer generation.}}

Finally, the \textit{Summarizer} serves as the synthesis point for all information flows. It receives inputs from two paths: direct, low-uncertainty cases from the \textit{Judger}, and comprehensively reasoned information from the \textit{Decoupler}. For the former, it performs a final verification. For the latter, it refines and integrates the global context and fine details to generate the final, faithful answer. Through this structured collaboration, our agentic generator dynamically adapts its reasoning depth, ensuring both efficient processing of simple queries and comprehensive understanding of complex ones.

\section{Experiments} 

\textbf{Data sources and settings.} 
As the first collection of open-domain visually rich documents, OpenDocVQA integrates multiple DocumentVQA datasets that span a broad range of real-world document types, as summarized in Table \ref{tab:table1}. Following the protocol in \cite{tanaka2025vdocrag}, we evaluate model performance on these public datasets under both supervised and zero-shot settings. To more faithfully reflect real-world usage scenarios, we further conduct evaluations under two retrieval configurations: single-pool and all-pool. In the single-pool setting, retrieval is restricted to the document pool corresponding to each individual dataset. In contrast, the all-pool setting requires models to retrieve from a unified, large-scale corpus drawn from diverse domains, providing a more realistic and challenging assessment of retrieval robustness and cross-domain generalization.

\textbf{Implementation details.} 
For fair comparison, we initialize our retriever with Phi3V \cite{Phi-3}, a state-of-the-art LVLM trained on high-resolution and multi-image datasets, consistent with prior works \cite{DSE,tanaka2025vdocrag}. We fine-tune the retriever using LoRA \cite{hu2022lora} under two configurations—with and without pre-training, while keeping the generator settings identical to those in \cite{tanaka2025vdocrag}. In HKRAG, the retriever and generator use independent parameter sets, and all other components remain frozen during fine-tuning. Training is conducted for one epoch using eight NVIDIA RTX 4090 GPUs, with the AdamW optimizer \cite{loshchilov2017decoupled}, FlashAttention \cite{dao2022flashattention} acceleration, and a batch size of 64. The temperature parameter $\tau$ is set to 0.01, submask number $\mathcal{N}$ is set to 2, and the Judger’s uncertainty threshold is fixed at 0.8.

\textbf{Unified evaluation metrics.} 
We evaluate retrieval performance using the nDCG@5 metric, a widely adopted standard in information retrieval \cite{ColPali,kamalloo2024resources}. For generation evaluation, we follow \cite{wang2025vidorag} and uniformly report accuracy. Specifically, we use the powerful Qwen-plus model to assess the quality of each prediction against its ground-truth answer, assigning a score on a 1–5 scale. A score of 4 indicates a generally correct but less concise response, while a score of 5 corresponds to a fully correct answer. Predictions receiving either score are treated as correct, and all others are counted as incorrect.

\subsection{Retrieval Results}

We evaluate HKRAG-Retriever against two categories of retrievers. The first category comprises off-the-shelf text retrieval models and image retrieval models, including BM25 \cite{BM25}, Contriver \cite{Contriever}, E5 \cite{E5}, GTE \cite{li2023GTE}, E5-Mistral \cite{E5Mistral}, NV-Embed-v2 \cite{NvEmbed2025}, CLIP \cite{CLIP2021}, DSE \cite{DSE}, and VisRAG-Ret \cite{VisRAG}. The second category consists of models fine-tuned on the OpenDocVQA dataset.

Table \ref{tab:table2} showcases the strong performance of HKRAG across four open-domain DocumentVQA benchmarks under both supervised and zero-shot settings. In the zero-shot setting on ChartQA and SlideVQA, HKRAG achieves new state-of-the-art results, with particularly notable gains in the challenging all-pool scenario (82.1\% on ChartQA and 72.5\% on SlideVQA). On the supervised benchmarks InfoVQA and DUDE, HKRAG also establishes new SOTA performance across all evaluation metrics, demonstrating excellent transferability and robustness.
These results confirm that HKRAG effectively captures query-relevant holistic knowledge from documents, substantially enhancing retrieval quality. Even when compared with fine-tuned vision–language models such as DSE, VisRAG-Ret, and VDocRAG, HKRAG consistently delivers superior performance. This advantage primarily stems from our efficient fine-tuning strategy, which enables precise localization of both salient knowledge and fine-grained knowledge.

\subsection{Retrieval-Augmented Generation Results}
As shown in Table \ref{tab:table3}, HKRAG-Generator consistently outperforms all competing methods across the four benchmarks. Under the All setting, it delivers clear performance improvements over VDocRAG—raising accuracy on ChartQA by 6.7\%, on SlideVQA by 3.5\%, on InfoVQA by 3.1\%, and on DUDE by 6.4\%. These gains stem from the complementary strengths of HKRAG-Retriever and HKRAG-Generator, which jointly enhance both document selection and holistic reasoning over visually rich content. Even in the Gold setting, HKRAG-Generator achieves up to 4.3\% higher accuracy than VDocGenerator, showing that its benefits are not merely due to improved retrieval but rather from more effective holistic-knowledge reasoning. Overall, the results demonstrate that HKRAG-Generator substantially alleviates the holistic-knowledge gap in open-domain visually rich document understanding, enabling more accurate and dependable answer generation.

\begin{table}[t]
\centering
\begin{small}
\renewcommand\arraystretch{1.1}{
\setlength{\tabcolsep}{3mm}{
\begin{tabular}{c|cc|cc}
\toprule
\multirow{2}{*}{\textbf{Method}} & \multicolumn{2}{c|}{\textbf{ChartQA}} & \multicolumn{2}{c}{\textbf{InfoVQA}} \\ \cline{2-5}
 & \textbf{Single} & \textbf{All} & \textbf{Single} & \textbf{All} \\ \hline
Baseline & 50.7 & 44.7 & 31.5* & 29.6* \\
HKRAG w/o Gen & 51.3 & 49.3 & 31.9* & 31.2* \\ 
HKRAG w/o Ret & 57.3 & 52.7 & 41.9* & 38.7* \\ 
HKRAG & \textbf{60.7} & \textbf{58.0} & \textbf{42.7}* & \textbf{39.6}*  \\ \hline 
\end{tabular}}}
\end{small}
\caption{Ablation study of Retriever (Ret) and Generator (Gen) in HKRAG. The baseline employs Eq. \ref{Eq1} for retriever fine-tuning, then directly inputs the generator.} 
\label{tab:table4}
\end{table}

\begin{table}[t]
\centering
\begin{small}
\renewcommand\arraystretch{1.1}{
\setlength{\tabcolsep}{3.2mm}{
\begin{tabular}{c|cc|cc}
\toprule
\multirow{2}{*}{\textbf{Method}} & \multicolumn{2}{c|}{\textbf{ChartQA}} & \multicolumn{2}{c}{\textbf{InfoVQA}} \\ \cline{2-5}
& \textbf{Single} & \textbf{All} & \textbf{Single} & \textbf{All} \\ \hline
Baseline & 84.4 & 76.3 & 70.4* & 53.9* \\
Ret w/o $\mathcal{L}_{\text{SIN}}$ & 84.9 & 77.9 & 71.3* & 55.2* \\
Ret w/o $\mathcal{L}_{\text{DIN}}$ & {85.7} & 79.4 & 70.7* & 54.6* \\
HKRAG-Ret & \textbf{87.6} & \textbf{82.1} & \textbf{72.4}* & \textbf{57.8}* \\ \hline 
\end{tabular}}}
\end{small}
\caption{Ablation study of different components in HKRAG-Ret.} 
\label{tab:table5}
\end{table}

\begin{table}[t]
\centering
\begin{small}
\renewcommand\arraystretch{1.1}{
\setlength{\tabcolsep}{2.2mm}{
\begin{tabular}{c|cc|cc}
\toprule
\multirow{2}{*}{\textbf{Method}} & \multicolumn{2}{c|}{\textbf{ChartQA}} & \multicolumn{2}{c}{\textbf{InfoVQA}} \\ \cline{2-5}
 & \textbf{Single} & \textbf{All} & \textbf{Single} & \textbf{All} \\ \hline
Baseline & 51.3 & 49.3 & 31.9* & 31.2* \\ 
Gen w/o pruner & 56.0 & 54.7 & 40.4* & 35.3* \\ 
Gen w/o decoupler & 59.0 & 56.3 & 40.5* & 38.0* \\ 
HKRAG-Gen & \textbf{60.7} & \textbf{58.0} & \textbf{42.7}* & \textbf{39.6}* \\ \hline 
\end{tabular}}}
\end{small}
\caption{Ablation study of different components in HKRAG-Gen. } 
\label{tab:table6}
\end{table}

\subsection{Analysis}
\textbf{Effect of each component in HKRAG.}
Table \ref{tab:table4} reports the ablation results of the hybrid masking–based holistic retriever and the uncertainty-guided agentic generator on ChartQA and InfoVQA. Removing either component results in clear performance drops, demonstrating that the two modules play complementary roles. The bidirectional retriever-tuning alone enhances retrieval accuracy by improving query–document alignment, whereas the generator alone yields a more substantial improvement in answer quality by adaptively selecting and reasoning over the informative documents. When both modules are enabled, HKRAG attains the highest performance across all settings, indicating that holistic retrieval and adaptive reasoning synergistically contribute to a more comprehensive understanding of visually rich documents.

\begin{figure}[t]
\centering
\begin{center}
{\includegraphics[width=0.478\textwidth]{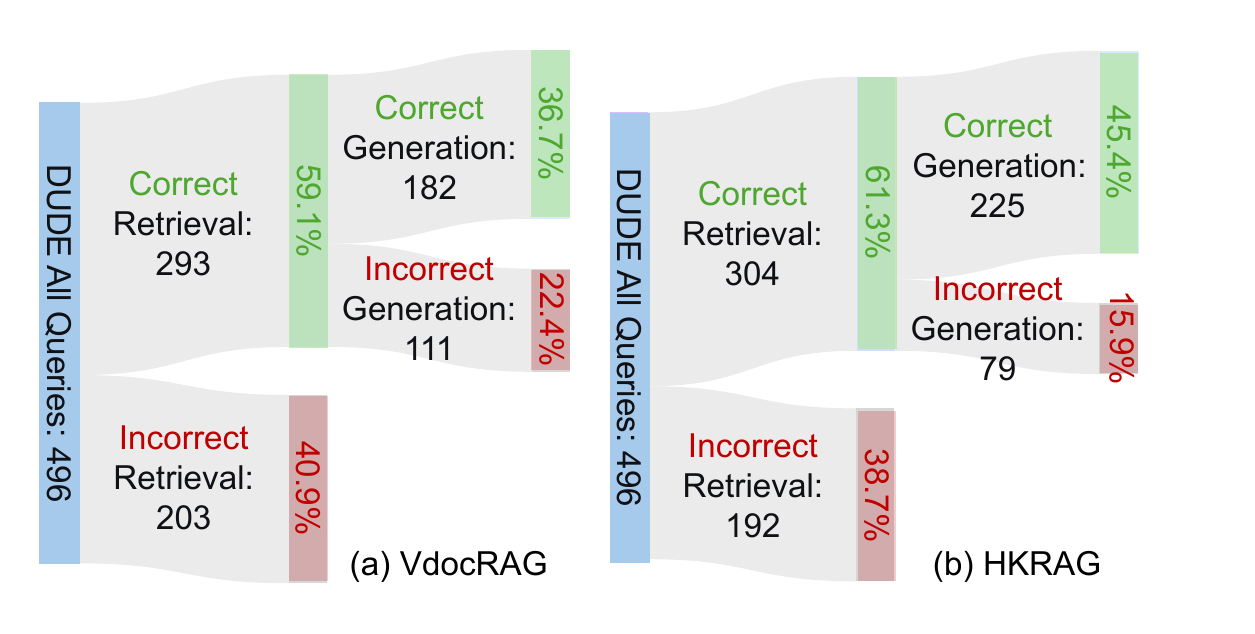}}
\caption{Performance of (a) VDocRAG and (b) HKRAG on DUDE. We present the distribution of queries across two categories: those that retrieved the correct document in the top-3 position (“correct retrieval”), and those that provided the correct answer given the top-3 retrieved documents (“correct generation”).} 
\label{fig:figure3.}
\end{center}
\end{figure}

\begin{figure}[t]
\centering
\begin{center}
{\includegraphics[width=0.37\textwidth]{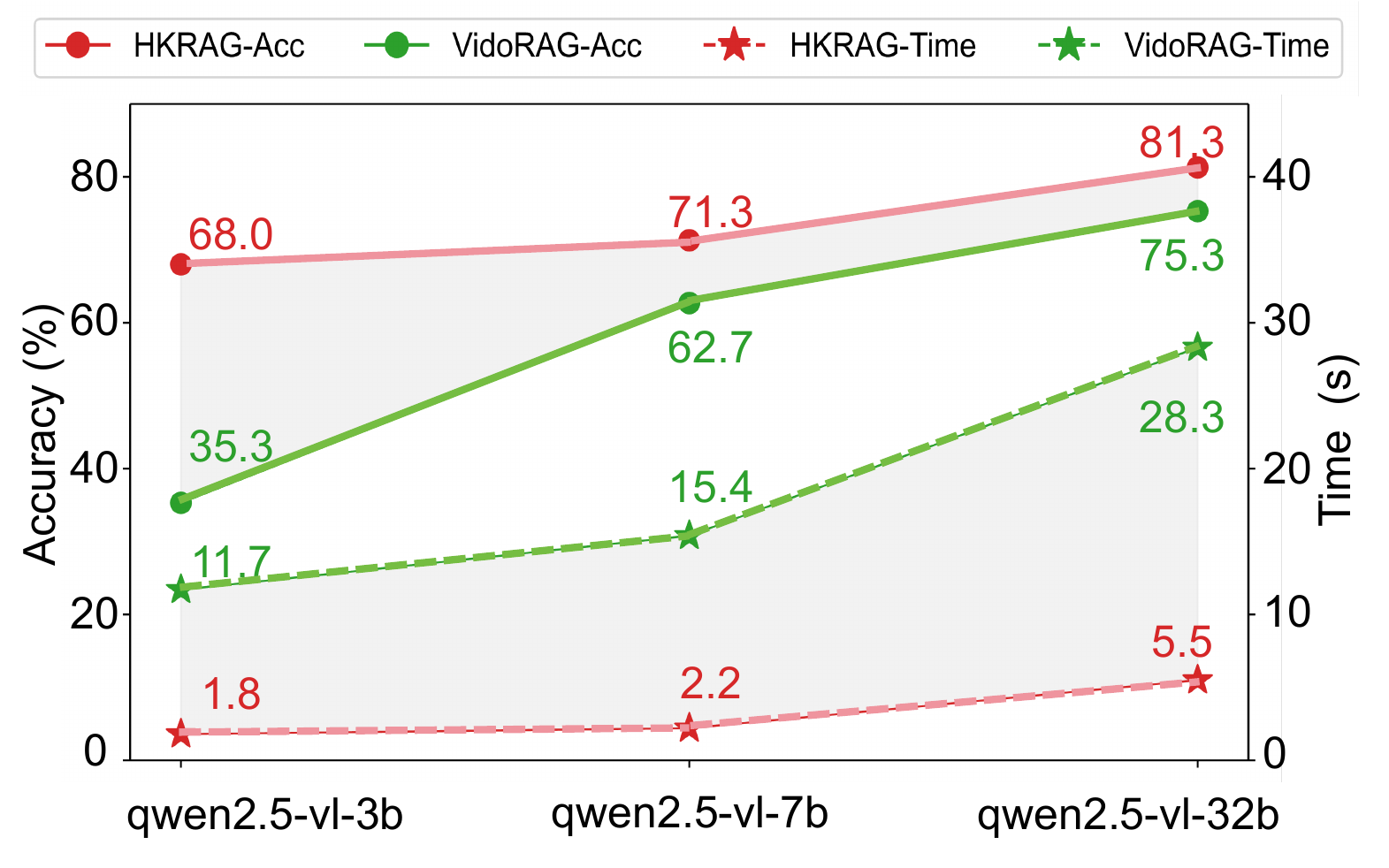}}
\caption{Comparison of VidoRAG \cite{wang2025vidorag} and our HKRAG across different sizes within the same series. The shaded area represents the gap between VidoRAG and HKRAG.}
\label{fig:figure4.}
\end{center}
\end{figure}

\begin{figure*}[ht]
\centering
\begin{center}
{\includegraphics[width=0.999\textwidth]{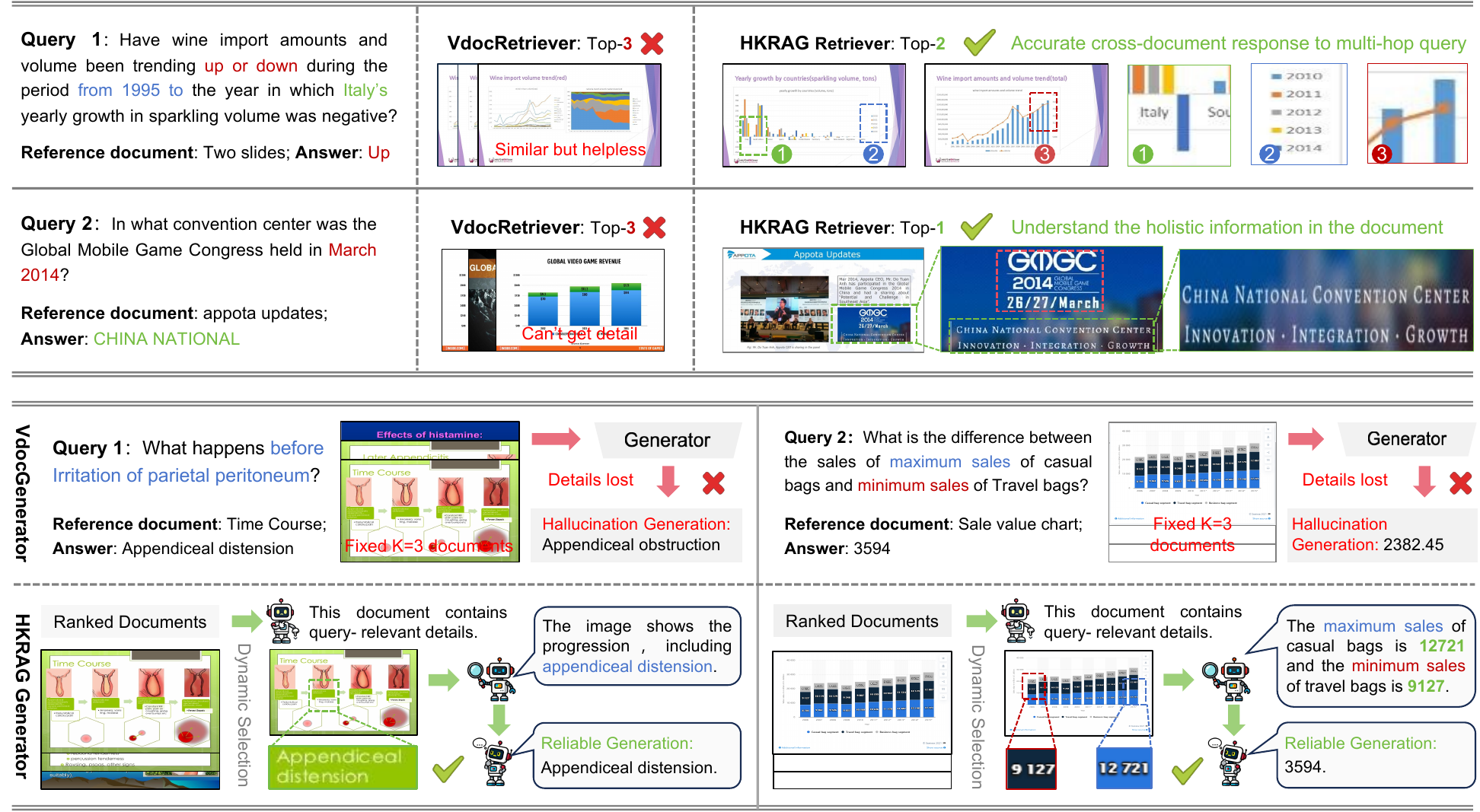}}
\caption{Qualitative results of our HKRAG compared to state-of-the-art VDocRAG \cite{tanaka2025vdocrag} on open-domain visually rich documents.} 
\label{fig:figure5.}
\end{center}
\end{figure*}

\textbf{How does our retriever gain benefits?}
Table \ref{tab:table5} presents the ablation results for the individual components of the Retriever. Removing either $\mathcal{L}{\text{SIN}}$ or $\mathcal{L}{\text{DIN}}$ leads to clear performance degradation, highlighting their complementary roles. Specifically, $\mathcal{L}{\text{SIN}}$ reinforces fine-print knowledge consistency, while $\mathcal{L}{\text{DIN}}$ strengthens salient knowledge consistency—both essential for reliable retrieval. The full HKRAG-Ret achieves the highest accuracy across all datasets and settings, demonstrating that jointly optimizing these objectives strikes an effective balance between fine-grained alignment and cross-domain generalization.

\textbf{How does our generator gain benefits?}
In Table \ref{tab:table6}, removing the dynamic buffer (Gen w/o pruner), which selects minimal sufficient knowledge documents instead of using fixed-K inputs, causes noticeable performance drops, particularly on InfoVQA, demonstrating its crucial role in efficient knowledge retrieval. Without the decoupler that handles HQP through reasoning disentanglement, performance degrades to 56.3 (ChartQA All) and 38.0 (InfoVQA All), indicating its importance in resolving ambiguous cases. HKRAG-Gen achieves the best results, validating that both components synergistically enhance knowledge selection and reasoning capability.

\textbf{How accuracy and time efficiency vary under different LVLM?}
In Figure \ref{fig:figure4.}, we demonstrate the accuracy and time efficiency of VidoRAG and HKRAG across different Qianwen 2.5 visual-language models, including 3b, 7b, and 32b. Through comparison, we observe that HKRAG consistently achieves higher accuracy than VidoRAG under identical settings while requiring less computational time. This indicates our generator efficiently and accurately infers answers, owing to our uncertainty-guided reasoning process that extracts holistic information based on uncertainty mining within query-document pairs.

\textbf{Qualitative results.}
We present a qualitative comparison with VDocRAG in Figure \ref{fig:figure5.}. During the retrieval phase (top), VDocRAG only includes documents semantically similar but irrelevant to Query 1 in its top-3 retrieval results. In contrast, our retriever accurately identifies two documents across the retrieval set that can answer the query, effectively addressing the multi-hop problem. During the generation phase (below), even when VDocRAG accurately retrieves documents containing answers, it generates hallucinations due to the difficulty of precisely inferring the computational relationship between maximum and minimum values from extensive visual information. Notably, our generator efficiently understands logical relationships within instances by dynamically selecting query-relevant documents amidst complex visual information.

\section{Conclusion} 
In this paper, we identify a critical limitation in existing multimodal RAG methods for open-domain DocumentVQA task: their inherent bias toward salient knowledge and consistent oversight of fine-print knowledge, which we term the \textit{inadequate holistic knowledge problem}. To address this, we propose HKRAG, a new multimodal RAG framework that enables holistic retrieval and dynamic integration of both knowledge types. Central to HKRAG are two new components: a hybrid masking-based holistic retriever, which explicitly enhances the model’s sensitivity to both salient and fine-print content through structured feature masking, and an uncertainty-guided agentic generator, which dynamically routes queries based on initial-answer confidence and performs iterative reasoning when needed. Extensive experiments on multiple DocumentVQA benchmarks, including zero-shot (ChartQA, SlideVQA) and supervised (InfoVQA, DUDE) settings, demonstrate that HKRAG consistently outperforms the compared baselines, confirming its robustness and generalizability in achieving truly holistic document understanding.

\small
\bibliography{sec/7_reference}
{
    \bibliographystyle{ieeenat_fullname}
}

\end{document}